\documentclass[aps,twocolumn,showpacs,floatfix]{revtex4}
\usepackage{graphicx}
\usepackage{epsfig}
\begin{document}
 
\title{Vacuum Energy and Repulsive Casimir Forces in Quantum Star Graphs}
 
\author{S. A. Fulling$^1$\footnote{Electronic address: fulling@math.tamu.edu;
\\URL: http://www.math.tamu.edu/$\sim$fulling},
L. Kaplan$^2$\footnote{Electronic address: lkaplan@tulane.edu;
\\URL: http://www.tulane.edu/$\sim$lkaplan},
and J. H. Wilson$^1$\footnote{Electronic address: thequark@tamu.edu}}

\affiliation{$^1$Departments of Mathematics and Physics, Texas A\&M 
University, 
  College Station, TX 77843-3368, USA \\
$^2$Department of Physics, Tulane University, New Orleans, LA 70118, USA
}

%\date{\today}
\date{March 25, 2007}

\begin{abstract}
 Casimir pistons are models in which finite Casimir forces can be 
calculated without any suspect renormalizations. It has been suggested 
that such forces are always attractive, but we present several 
counterexamples, notably a simple type of quantum graph in which the 
sign of the force depends upon the number of edges. We also show that 
Casimir forces in quantum graphs can be reliably computed by summing over the classical 
orbits, and study the rate of convergence of the periodic orbit expansion. 
In generic situations where no analytic expression is available, the sign 
and approximate magnitude of Casimir forces can often be obtained using 
only the shortest classical orbits. \end{abstract}


\maketitle 

\section{Introduction}

\begin{figure}
\begin{picture}(200,60)
 \put(0,0){\framebox(200,60){}}
 \put(40,0){\line(0,1){60}}
 \put(2,30){$b_1$}
 \put(20,2){$a$}
 \put(120,2){$L-a$}  
\end{picture}
 \caption{A rectangular piston in 
two dimensions (cf.~\cite{Cav}). 
 In three dimensions   there is another length, $b_2\,$, perpendicular 
to the plane of the figure.} 
 \label{fig:piston}
 \end{figure}
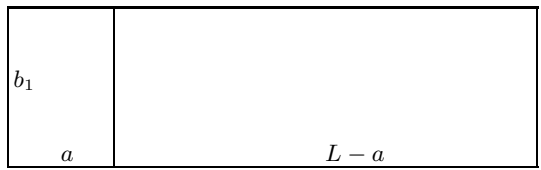

  According to a classic calculation \cite{Luk}, the Casimir force 
  inside a roughly cubical rectangular parallelepiped is repulsive;
 that is, it tends to expand the box.  
The reasoning leading to this conclusion is open to criticism on 
two related grounds: It ignores the possibility of nontrivial 
vacuum energy in the region outside the box, and it involves 
``renormalization'' in the sense of discarding divergent terms 
associated with the boundary although (unlike the case of 
parallel plates, or any calculation of forces between rigid bodies)
 the geometry of the boundary depends upon the dimensions of the 
box.
 Recently 
 (see also \cite{SS})
 a class of scenarios called ``Casimir pistons'' has been 
introduced to which these objections do not apply.
 The  piston is an idealized plate that is free to move along a 
rectangular shaft, whose length, $L-a$,
  to the right of the piston is taken 
arbitrarily large (Fig.~\ref{fig:piston}).
 Both the external region and the divergent
 (or cutoff-dependent) terms in the internal 
vacuum energy are independent of the piston position, $a$,
 so that a well-defined, finite force on the piston is calculated.
 One finds that this force is always attractive,  both for
  a two-dimensional scalar-field model with the Dirichlet boundary 
  condition~\cite{Cav} and for a three-dimensional 
 electromagnetic field with the perfect-conductor 
boundary condition~\cite{HJKS}. 

 Barton \cite{Bar} showed that the piston force can be repulsive 
for some (not too small) values of~$a$ if the conducting material 
is replaced by a weakly polarizable dielectric. 
This result is somewhat ironic in that one reason for suspicion of 
repulsive Casimir forces is the belief that the force between 
disjoint bodies of realistically modeled material should be always 
attractive.
The unexpected result is easily understood, however, as being due to attraction 
between the piston and the distant part of the shaft.
The effect would disappear if the shaft extended a long distance to the 
left of the fixed plate (``baffle'') at $a=0$ as well as to the 
right of the piston.

 In the present paper we study the vacuum energy and Casimir forces in 
one-dimensional quantum graph models and observe several 
situations with idealized boundary conditions for which the piston force 
is unambiguously repulsive. In quantum graphs of high symmetry, the 
Casimir forces may be calculated analytically. More generally, we show 
that these forces may be obtained systematically from a sum over the 
classical periodic orbits in the graph,
as done in three-dimensional problems in \cite{JR,perorb,JS}, and we 
discuss the 
rate of convergence of the periodic-orbit expansion. In some cases, the 
sign and approximate magnitude of the force on a Casimir piston may be 
obtained using only the shortest orbit hitting that piston. Although the 
quantum graph models are less realistic than those studied in \cite{HJKS} 
and~\cite{Bar}, they do show that repulsive Casimir forces do arise 
physically and are not inevitably an artifact of a naive renormalization 
scheme. Our effects are unrelated to that in~\cite{Bar} and do not depend 
on the asymmetry noted above in connection with that paper. 
The periodic-orbit techniques discussed here  have relevance to the 
study of 
Casimir energies in more realistic geometries  (cf.~\cite{schad}),
 including two- and three-dimensional chaotic billiards. 
In an Appendix, we consider a 
situation in which an unambiguously repulsive Casimir force appears for 
the electromagnetic field in a three-dimensional geometry.

Throughout, we take $\hbar=1=c$.

\section{Vacuum Energy in Quantum Graphs}
 
A finite quantum graph~\cite{Roth-1983,kottossmil,Kuc,gnutzmann} consists 
of $B$ one-dimensional 
undirected bonds or edges of length $L_j$ ($j=1, \ldots , B$). Either end 
of each bond ends at one of $V$ vertices, and the valence $v_\alpha \ge 1$ 
of a vertex is defined as the number of bonds meeting there. A normal mode 
$u$ of the quantum graph has the form $u_j(x)=a_j \cos(k x_j)+b_j \sin(k 
x_j)$ on every bond $j$, and satisfies the specified boundary conditions 
at each vertex. Despite their simplicity, quantum graph models have 
previously shed light on a number of important physical problems, having 
served originally as models of conjugated molecules, and more recently of 
quantum, electromagnetic, and acoustic waveguides and circuits. These 
models have also served as valuable testing grounds for studying more 
general properties of quantum behavior, including Anderson localization, 
quantum chaos, adiabatic quantum transport, and scattering. A recent 
review may be found in~\cite{Kuc2002}.

In the spirit of abstract modeling, the vacuum energy of a graph is 
defined as  
the sum (renormalized) of zero-point energies 
over all normal-mode frequencies $\omega_n\,$, where the frequency 
$\omega_n$ is equal to the wave number $k_n$ in our units. It is 
convenient to apply an exponential ultraviolet regularization (the same 
answer would be obtained, for example, by a calculation with zeta 
functions):
\begin{equation}
 E(t) \equiv \sum_{n=0}^\infty {1 \over 2} \omega_n e^{-\omega_n t}
 = -{1 \over 2} {d \over dt} T(t)\,,
\label{vacenergy}\end{equation}
where 
 \begin{equation}
 T(t) \equiv \sum_{n=0}^\infty e^{-\omega_n t}
 \label{cyltrace}\end{equation}
is the trace of the so-called cylinder kernel \cite{systematics}.

\section{Analytic Examples of Repulsive Casimir Forces}

\subsection{One-dimensional piston with mixed boundary conditions}
The first example is already rather well known, in its essence.
 Consider a scalar field quantized on a line divided into 
three parts by two points, at each of which either a Dirichlet or a 
Neumann boundary condition is imposed.
 The contributions of the two infinite 
 (or, better, extremely long) intervals 
to the Casimir force will vanish.
 (As emphasized in~\cite{HJKS}, the force contributed by a long 
shaft is entirely associated with periodic orbits perpendicular to 
the shaft, which do not exist in the one-dimensional case.)
 Let the length of the central interval be~$a$.
 Then the frequencies of the normal modes are 
 \begin{equation}
 \omega_n = \frac{n\pi}{a} 
 \label{DD} \end{equation}
 for nonnegative (or positive) integer~$n$,
 if the boundaries are both Neumann (or both Dirichlet, 
respectively),
and one has
 \begin{eqnarray}
 T(t) &=&  \sum_{n=0,1}^\infty e^{-\pi nt/a} \nonumber\\
 &=& \frac{1}{1 -e^{-\pi t/a} } \quad[{}-1] \\
 &=& \frac a{\pi t} \pm \frac{1}{2} + \frac1{12}
 \,\frac{\pi t}{a} +O(t^2) \nonumber \,.
 \end{eqnarray}
Thus the regularized vacuum energy is
 \begin{equation}
 E(t) = \frac  a{2\pi t^2} -\frac{\pi}{24a} + O(t) \,.
 \label{DDenergy}\end{equation}
The leading, divergent term is proportional to the interval length $a$ and corresponds to
a geometry-independent constant energy density. This term is compensated 
 in the force
 by similar terms in the exterior regions, already discarded. Then 
letting $t\to0$, we obtain the well-known attractive force
 \begin{equation}
 F \equiv -\, \frac{\partial E}{\partial a} = -\,\frac{\pi}{24a^2}\,.
 \label{DDforce}  \end{equation}

 More precisely, if the entire space has length~$L$, then the 
regularized energy of the exterior regions is 
\begin{equation} \frac{L-a}{2\pi t^2} + O(L^{-1}) \,. \end{equation}
 The second term is negligible as $L\to\infty$, and the first term 
combines with the first term of (\ref{DDenergy}) to make a term 
independent of~$a$, which, therefore, is an unobservable constant energy shift
that contributes nothing to the 
force.
 Henceforth we shall not repeat this type of argument every time it 
is needed, and will simply refer to such endpoints  as 
Neumann or Dirichlet pistons.

On the other hand, if one boundary is Dirichlet and the other 
Neumann, then the eigenfrequencies are
 \begin{equation}
 \omega_n = \frac{(2n+1)\pi}{2a} \,.
 \label{DN} \end{equation}
The same calculation leads to
\begin{eqnarray}
 T(t) &=& e^{-\pi t/2a} \sum_{n=0}^\infty e^{-\pi nt/a} \nonumber \\
 &=& \frac1{2\sinh(\pi t /2a)} \\
 &=&
  \frac a{\pi t} - \frac1{24} \,\frac{\pi t}{a} +O(t^2); \nonumber
 \end{eqnarray}
the regularized energy is
 \begin{equation}
 E(t) =   \frac  a{2\pi t^2} +\frac{\pi}{48a} + O(t)\,,
 \label{DNenergy}\end{equation}
  and the force comes out to be repulsive:
 \begin{equation}
 F  = +\,\frac{\pi}{48a^2}\,.
 \label{DNforce}  \end{equation}
 
 \subsection{Quantum star graphs}

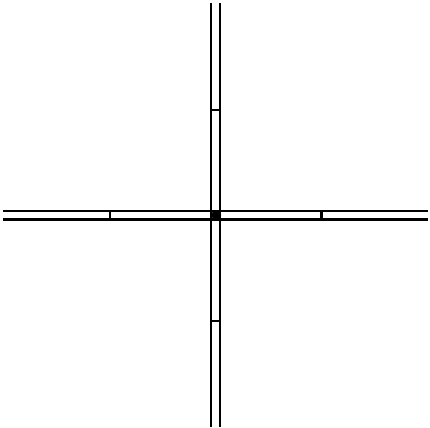
\begin{figure}
 \begin{picture}(160,160)(-80,-80)
 \put(-2,-2){$\bullet$}
 \put(1.6,0){\line(0,1){80}}
  \put(-1.6,0){\line(0,1){80}}
\put(-1.6,40){\line(1,0){3.2}}
 \put(1.6,0){\line(0,-1){80}}
  \put(-1.6,0){\line(0,-1){80}}
\put(-1.6,-40){\line(1,0){3.2}}
 \put(0,1.6){\line(1,0){80}}
  \put(0,-1.6){\line(1,0){80}}
\put(40,-1.6){\line(0,1){3.2}}
 \put(0,1.6){\line(-1,0){80}}
  \put(0,-1.6){\line(-1,0){80}}
\put(-40,1.6){\line(0,-1){3.2}}
\end{picture}
 \caption{A star graph with a piston installed in each edge.
 (The pistons are actually 
points; the edges have no thickness.)}
 \label{fig:star}
 \end{figure}

 In the next model the space consists of $B$ one-dimensional rays of 
large length $L$ attached to a central vertex (Fig.~\ref{fig:star}).
 In each ray a Neumann piston is located a distance $a$ from the 
vertex, so that a normal mode of the field in ray~$j$ must take the 
form $u_j(x) = c_j \cos\bigl(\omega(x-a)\bigr)$ when $x$ is 
measured from the center.
 At the central vertex the field has the Kirchhoff 
(generalized Neumann) behavior 
 \begin{equation}
 u_j(0) =C  \hbox{ for all $j$} , \quad
 \sum_{j=1}^B u'_j(0)=0 \,.
 \label{Kirch}\end{equation}
The following analysis is part of a broader study of vacuum energy 
in quantum graphs~\cite{Wil} (see also \cite{Fsb,BM,BHW}).

 There are two types of normal modes.
 First, if $\,\cos(\omega a) \ne 0$, we have from (\ref{Kirch})
that  $c_j= C/\cos(\omega a)$ and $\,\tan(\omega a)=0$,
 whence $\omega$ is one of the numbers~(\ref{DD}).
 Second, if $\,\cos(\omega a) = 0$, then $\omega$ 
 is one of the numbers~(\ref{DN}) and 
 \begin{equation} \sum_{j=1}^B c_j=0 \,, \end{equation}
 which has $B-1$ independent solutions.
 Therefore, the energies and forces are just the appropriate linear 
combinations of those calculated in the previous example:
 the regularized energy for the whole system is
 \begin{equation}
 E(t) = \frac{BL}{2\pi t^2} +\frac{(B-3)\pi}{48a}
 +O(L^{-1}) + O(t) \,, 
 \label{starenergy}\end{equation}
 and the force (either from (\ref{starenergy}) or from 
(\ref{DDforce}) and (\ref{DNforce})) is
 \begin{equation}
 F= -\frac{\pi}{24a^2} + (B-1)\frac{\pi}{48a^2}
 =\frac{(B-3)\pi}{48a^2}\,.
 \label{starforce}\end{equation}

 When $B=1$ or $B=2$, the result reduces properly to that for an 
ordinary Neumann interval of length $a$ or $2a$, respectively.
 When $B>3$, however, the force is repulsive:
if the pistons are free to all move together, they will tend to move outward.
 (More generally, a periodic-orbit calculation, such as discussed
 in  Section~\ref{secorbit}, is applicable to 
 unequal piston displacements and confirms that the force on each 
individual piston is outward, so there are no other, asymmetrical 
 modes that are partly attractive.)
 This repulsive effect cannot be attributed to mixed boundary conditions, 
since all the conditions are of the Neumann type.
 (However, replacing all the pistons with Dirichlet pistons while 
maintaining (\ref{Kirch}) would interchange the roles of the two 
types of eigenvalues and produce attraction for all $B>1$.)

\section{Periodic-Orbit Calculations for General Graphs}
\label{secorbit}

For a general quantum graph, e.g., for a star graph with unequal bond
lengths or with more complicated boundary conditions,
no simple expressions for the normal-mode frequencies $\omega_n$
are available,
and thus the vacuum energy and Casimir forces cannot
be computed in closed form. Computing the spectrum numerically,
as discussed below, allows for an accurate evaluation of the
vacuum energy for any specific quantum graph, but this type of brute
force calculation must be repeated anew for every geometry and does
not provide much physical insight regarding the attractive or repulsive
character of Casimir forces in different cases. Instead, much intuition
may be obtained using a classical-orbit approach, where the sign and
magnitude of every contribution to the vacuum energy are seen to be
directly related to bond lengths and boundary conditions at the vertices.

It is convenient to describe boundary conditions at every vertex $\alpha$
by a unitary $v_\alpha \times v_\alpha$ scattering matrix $\sigma_\alpha$
(which acts on the space of {\it undirected} bonds meeting at vertex $\alpha$).
For example, a Neumann or Dirichlet boundary condition at a vertex of valence
$v_\alpha=1$ corresponds to a scattering matrix $\sigma_\alpha=(+1)$ or $(-1)$,
respectively, while the Kirchhoff boundary condition is described by
$(\sigma_\alpha)_{jj'}={2 \over v_\alpha} -\delta_{jj'}$. Together these
constitute a $2B \times 2B$ scattering matrix $S$ for the entire graph of
$2B$ {\it directed} bonds~\cite{kottossmil,gnutzmann}
or bond-ends~\cite{kswire,ksinverse,kostrykin-2007}. 
To make the following arguments valid,
we must assume that $S$ is independent of energy or frequency
($k$-independent), 
as is true for the Dirichlet,
Neumann, and Kirchhoff boundary conditions we treat here 
(but not for the more
general Kirchhoff-type boundary conditions where a potential is attached to
 each vertex (\cite{ES} and \cite{kottossmil,Kuc,Fsb})). 
Then one can construct \cite{kottossmil,gnutzmann} 
  a trace formula relating
the spectrum of a graph 
(away from the point $\omega=0$, which makes no contribution to vacuum 
energy anyway)
to its periodic orbits,
\begin{equation}
\label{PrimTrace}
\sum_n \delta(\omega-\omega_n) = \frac{L}{\pi} 
+ {\rm Re}\frac{1}{\pi} \sum_p
\sum_{r=1}^\infty  (A_p)^r {L_p} e^{ir \omega L_p} \,.
\end{equation}
(Variations on the trace formula have been found in \cite{Roth-1983},
\cite{kostrykin-2007}, \cite{Wil}, and elsewhere.)
In (\ref{PrimTrace}) the values $\omega_n$ are the normal-mode 
frequencies,
$L=\sum_{j=1}^B L_j$ is the total length of the graph, which determines
the smooth (Weyl) contribution to the spectrum, and the sum over $p$ is a
sum over primitive periodic orbits (orbits that cannot be written as
repetitions of shorter orbits). Each $p$ takes the form
$p = j_1 j_2 \cdots j_n$ where every $j_i$ is a directed bond. The 
corresponding amplitude of the primitive periodic orbit is given by a
product of scattering factors,
$A_p = S_{j_1 j_2} \cdots S_{j_{n-1} j_n} S_{j_n j_1}$, the metric length
of the primitive orbit is $L_p=L_{j_1}+\cdots+L_{j_n}$, and each $r$ is a
different repetition number of our base primitive orbit. 

Substituting the spectrum given by Eq.~(\ref{PrimTrace}) into
Eq.~(\ref{vacenergy}), we obtain
\begin{equation}
\label{Energy}
E(t) = \frac{L}{2 \pi t^2} - {\rm Re} \frac{1}{2\pi} \sum_{p}
\sum_{r=1}^\infty \frac{(A_p)^r}{L_p r^2} + O(t) \,.
\end{equation}
As discussed previously, the finite vacuum energy, which is relevant
for computation of Casimir forces,
is obtained by dropping the divergent Weyl term and taking the limit $t \to 0$,
\begin{equation}
\label{VacEner}
E_c = -\, \frac{1}{2\pi} {\rm Re} \sum_p \sum_{r=1}^\infty 
\frac{(A_p)^r}{L_p r^2} \,.
\end{equation}
A mathematically rigorous derivation and proof of (conditional)
convergence of Eq.~(\ref{VacEner}) will appear in~\cite{BHW}.

Equivalently, we may begin with the free cylinder kernel in one dimension,
\begin{equation}
T_0(x,x',t) = \frac{t}{\pi} \frac{1}{(x-x')^2+t^2} \,,
\end{equation}
apply the method of images to include scattering from the vertices,
take the trace
\begin{eqnarray} \label{ttrace}
T(t)&=&\int dx \, T(x,x,t) \nonumber \\ &=& \frac{t}{\pi}\frac{L}{t^2} +
{\rm Re} \sum_p \sum_{r=1}^\infty \frac{t}{\pi}  \frac{2L_p (A_p)^r}{(r L_p)^2}
+O(t^2)
\,, \end{eqnarray}
and finally
use Eq.~(\ref{vacenergy}) to obtain the result (\ref{Energy}).
(This construction, which generalizes the study of the heat 
kernel in \cite{Roth-1983}, is described in detail in \cite{Wil}.) 

The Casimir force on any piston may be obtained easily by differentiating
Eq.~(\ref{VacEner}) term by term with respect to the appropriate bond length
$L_j$\,.

 We note that the expansion (\ref{VacEner}) of the vacuum energy
is exact and involves periodic orbits only.
 The derivation of Eq.~(\ref{ttrace})
 hinges on the identity $(\sigma_\alpha)^2=I$
for the scattering matrix at each vertex.
 This condition holds for any $k$-independent
scattering matrix~\cite{kswire,ksinverse}, including real scattering 
matrices
of the form used here, but also complex energy-independent scattering
matrices in the case of time-reversal symmetry breaking by magnetic 
fields.  It is the crucial ingredient in proving that closed but 
nonperiodic paths (i.e., paths that start and end at~$x$ but with opposite 
momenta)
make no net contribution to the vacuum energy.
 When $S$ depends on~$k$, two complications arise.
First,  the method of images cannot be so easily applied to  
``time-domain'' integral kernels such as $T$ and the heat kernel,
because the reflection law becomes nonlocal in~$t$.
Second,   the identity $(\sigma_\alpha)^2=I$
no longer applies, and the nonperiodic paths make a nontrivial 
contribution to the vacuum energy
(and to the density of states, Eq.~(\ref{PrimTrace}), even when 
$\omega\ne0$).
Both effects are visible in the investigations of the simplest special 
cases in \cite{BF,Fsb}.

To evaluate the accuracy of the periodic-orbit expansion in situations
where no analytic expression for the vacuum energy is available, we may
compare with a brute-force calculation where the spectrum is evaluated
numerically. For a general $V$-vertex graph, the normal-mode frequencies
are given by solutions of a characteristic equation 
$\,{\rm det} \; h(\omega)=0$,
where $h(\omega)$ is a $V\times V$ matrix~\cite{kottossmil}. For the special case of a star graph
with irrationally related bond lengths, we have
\begin{equation}
\sum_{j=1}^B \tan(\omega L_j + \theta_j) = 0 \,,
\end{equation}
where $\theta_j=0$ or $\pi$ for a Neumann or Dirichlet piston on 
bond~$j$, respectively.
In any case, given a method for obtaining a numerical spectrum 
$\omega_n\,$,
 we may evaluate
\begin{equation}
E_{\rm finite}(t)=\sum_n {1 \over 2} \omega_n e^{-\omega_n t}
 -\frac{L}{2\pi t^2}
\label{vacnumer}\end{equation}
to any desired accuracy by summing over all $\omega_n \le \omega_{\rm max}$
where $\omega_{\rm max} \gg 1/t$. Since the divergent term associated with
the Weyl density of states, or equivalently with the free one-dimensional
geometry, has already been subtracted, we only need take the numerical limit
$t \to 0$ to obtain the true vacuum energy $E_c\,$. Expressing the 
regularized
vacuum energy as a power series,
\begin{equation}
E_{\rm finite}(t) =E_c + \alpha_1 t +\alpha_2 t^2 +\cdots \,,
\end{equation}
we may apply Richardson extrapolation to approximate the vacuum energy
to any desired order of accuracy, $E_c=E_c^{\rm numerical}+O(t^s)$, by
evaluating $E_{\rm finite}(t)$ at $s$ distinct values of the regularization
parameter $t$.

\section{Rate of Convergence of Periodic-Orbit Expansion}
\label{secconverg}

We consider a star graph with Kirchhoff boundary condition for $B$
bonds meeting at the central vertex, and a Dirichlet or Neumann piston
on each bond at a distance $a_j$ from the central vertex (i.e., the pistons
may be located at different distances from the center). The leading
contribution to the vacuum energy is given by the shortest primitive
orbits, each of which travels back and forth along a single bond.
Including all repetitions of such orbits, we obtain
\begin{equation}
E_c^{\rm shortest} = -\frac{1}{4 \pi} \sum_{j=1}^B \sum_{r=1}^\infty
{1 \over r^2}\left(\frac{2}{B}-1\right)^r \frac{\cos(r\theta_j)}{a_j} \,,
\label{shortest}
\end{equation}
where $\theta_j=0$ for a Neumann piston or $\pi$ for a Dirichlet
piston. For example, for all Neumann pistons the 
sum over $r$ can be evaluated as a dilogarithm, which in turn can
be expanded in powers of $1/B$ as
\begin{equation}
E_c^{\rm shortest} = \frac{\pi}{48} \left(1-\frac{24 \ln 2}{\pi^2B}
+\cdots\right) \sum_{j=1}^B \frac{1}{a_j} \,.
\end{equation}
This approximation compares well to  the analytic result
$\frac{\pi}{48}\left(1-\frac{3}{B}\right)\frac{B}{a}$ for
$B$ {\it equal-length} bonds (Eq.~(\ref{starenergy})).

\begin{figure}
\centerline{\includegraphics[width=3.3in,angle=0]{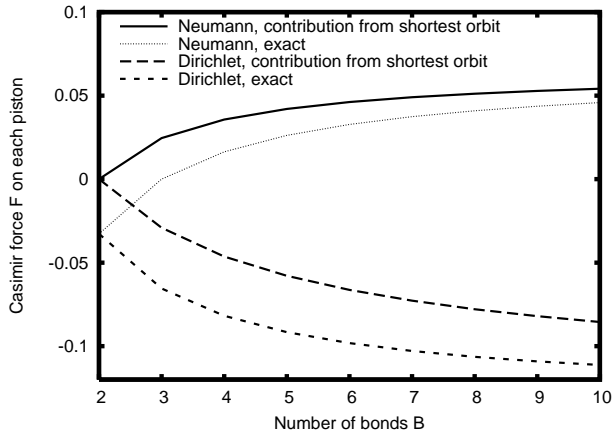}}
\protect\caption{The force on a piston in a star graph with $B$ bonds
of length 1, Kirchhoff boundary condition at the center, and either
Neumann or Dirichlet boundary condition at each piston is computed
using only the shortest periodic orbit (Eq.~(\ref{shortest})) and
compared with the exact answer. Positive values indicate repulsive forces.}
\label{figshort}
\end{figure}

The results are illustrated in Fig.~\ref{figshort}, where the exact
force on each piston in a star graph having either all Dirichlet or
all Neumann pistons is compared with the contribution to the force
from the shortest periodic orbit. The repulsive behavior in the Neumann
case, as well as the attractive behavior in the Dirichlet case, are well
explained by considering only the shortest periodic orbit, i.e., the
bounce between the piston and the central vertex.

\begin{figure}
\centerline{\includegraphics[width=3.3in,angle=0]{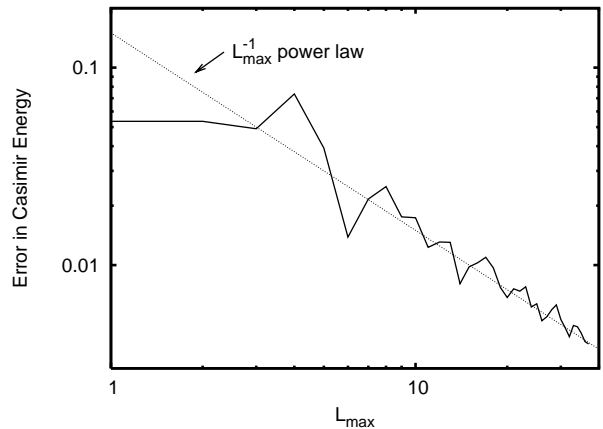}}
\protect\caption{The error $|E_c^{L_{\rm max}}-E_c|$ in the periodic
orbit expansion for the vacuum energy is shown for a star graph with
four bonds of length $1.1$, $1.6176$, $1.2985$, and $1.1159$, and a
Neumann piston at the end of each bond.}
\label{figconverg1}
\end{figure}

To obtain a better approximation, we may systematically include
contributions from longer orbits. In Fig.~\ref{figconverg1}, we
show the convergence of the sum (\ref{VacEner}) when all orbits,
including primitive orbits and repetitions, of total length
$rL_p \le L_{\rm max}$ are included in the summation. In this example,
a star graph with $B=4$ bonds, all Neumann pistons, and unequal bond
lengths is used, so the exact answer is obtained to the necessary accuracy
from a numerical spectrum as described in Section~\ref{secorbit}. We note
that the rate of convergence is given by
\begin{equation}
|E_c^{L_{\rm max}}-E_c| \sim \frac{1}{L_{\rm max}} \,,
\end{equation}
consistent with the fact that each contribution to Eq.~(\ref{VacEner})
from orbits of length $rL_p \in [L_{\rm max},L_{\rm max}+\Delta]$ scales
as $L_{\rm max}^{-2}$ for large $L_{\rm max}$, and all such contributions
appear preferentially with the same (negative) sign.

\begin{figure}
\centerline{\includegraphics[width=3.3in,angle=0]{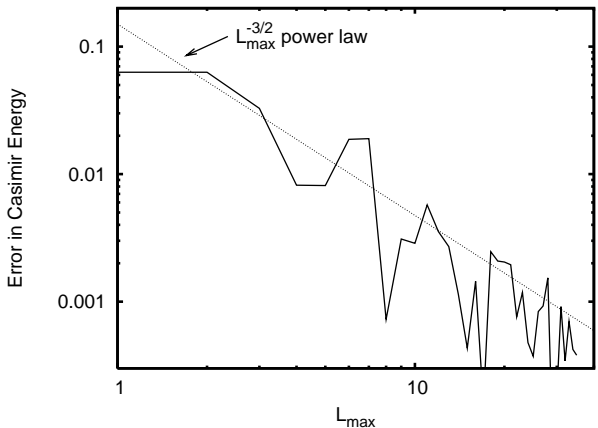}}
\protect\caption{The error $|E_c^{L_{\rm max}}-E_c|$ is shown for the
same quantum graph as in Fig.~\ref{figconverg1}, but with a Dirichlet
piston at the end of the first bond.}
\label{figconverg2}
\end{figure}

In more general situations, involving non-star topologies, more complicated
boundary conditions, or non-zero gauge fields, orbits of different length
are expected to contribute with random signs to the sum (\ref{VacEner}).
The error made by omitting orbits of length greater than $L_{\rm max}$
takes the form $\sum_{n=0}^\infty D_n\,$, where $D_n\,$, associated with 
all
orbits of total length $rL_p \in [L_{\rm max}+n\Delta,
L_{\rm max}+(n+1)\Delta]$, scales as $D_n \sim (L_{\rm 
max}+n\Delta)^{-2}$,
but the $D_n$ appear with random (uncorrelated) signs. The mean squared
error then scales as $\sum_{n=0}^\infty D_n^2 \sim
\sum_{n=0}^\infty (L_{\rm max}+n\Delta)^{-4} \sim L_{\rm max}^{-3}\,$,
and the root mean square error decays as
\begin{equation}
|E_c^{L_{\rm max}}-E_c| \sim \frac{1}{L_{\rm max}^{3/2}} \,.
\label{decay2}
\end{equation}
As an example, in Fig.~\ref{figconverg2}, we consider the convergence
of the periodic-orbit sum for the same $4$-bond star graph, but with
a Dirichlet instead of Neumann piston on one of the bonds. The behavior
is consistent with the faster rate of convergence predicted by 
Eq.~(\ref{decay2}).

\section{Summary}

We have seen that unambiguously repulsive as well as unambiguously 
attractive Casimir forces arise in simple quantum-graph models, and that 
the sign of the force in a given geometry may often be easily understood 
in terms of the short periodic orbits of the system. We have also examined 
(numerically)
the rate of convergence of the periodic-orbit expansion. Classical-orbit 
approximations may also be useful 
for understanding the sign of Casimir forces in higher-dimensional piston 
systems where no analytic solution exists, for example, in two- or 
three-dimensional chaotic billiards.

\section*{Acknowledgments}
        We thank Kimball Milton and the Texas A\&M 
quantum graph research group (Brian Winn, Gregory Berkolaiko, and 
Jonathan Harrison) for helpful comments.  This research is 
supported in part by National Science Foundation Grants
PHY-0554849 and PHY-0545390.
We thank the Isaac Newton Institute for Mathematical Sciences in 
Cambridge, U.K., for hosting an extended, supported visit by S.A.F. and a 
brief visit by J.H.W.

\appendix \section{Infinitely permeable piston}
\label{secappendix}
 In principle, a repulsive piston can be constructed in the more 
realistic case of the electromagnetic field in dimension~3,
 in analogy with our original one-dimensional model.
 If the electromagnetic analog of the Dirichlet condition is a 
perfect conductor, then the analog of the Neumann condition is a 
material with infinite magnetic permeability, and the Casimir force 
between slabs of these two different types is repulsive~\cite{Boy}.
 (A list of references on this topic appears in \cite{AFG}.)
 The existence of real materials with sufficient permeability to 
exhibit Casimir repulsion in the laboratory is controversial
 \cite{IC,KKMRrep,SZL}.
 Here we merely check that the piston effect discovered by 
Cavalcanti \cite{Cav} and the MIT group \cite{HJKS} does not 
destroy the repulsion shown by less sophisticated calculations.
This is not trivial, since the effect arises from the action of the 
shaft walls on the transverse behavior of the field.

 Following Lukosz~\cite{Luk}, but in a notation closer to 
Cavalcanti's (see Fig.~\ref{fig:piston}), we consider a rectangular 
box with dimensions $a$, $b_1\,$, and $b_2\,$.
 As previously exemplified, we can calculate a finite vacuum energy 
naively, in full confidence that the discarded divergent terms will 
cancel when a force is calculated for the piston system as a whole.
 We are interested in the case where the piston (the surface that 
is free to move) is infinitely permeable but the shaft and the 
baffle (the rest of the box) are perfect conductors.
 By the Rayleigh--Dowker argument~\cite{Dow},
 the energy, $\overline{E}_a\,$, of such a box is
 \begin{equation}
 \overline{E}_a = E_{2a} - E_a\,,
 \label{Rayleigh}\end{equation}
 where $E_a$ is the energy of a totally conducting box also of 
length~$a$.
 By differentiation with respect to~$a$ (not $2a\,$!), this 
relation extends to forces and pressures.
(Throughout this discussion ``pressure'' simply means ``force 
per area'' without necessarily implying a local pressure 
independent of position on the wall.) 
 Thus (\ref{DNforce}) follows from (\ref{DDforce}) by virtue of
 \begin{equation} -\frac{\pi} {24a} \left[ \frac12 -1\right] = 
 -\frac{\pi} {24a}  \left[-\,\frac12\right], \end{equation}
 and the three-dimensional analogs will involve quantities 
proportional to
\begin{equation} \frac1{a^3} \left[\frac18 - 1\right] = 
 \frac1{a^3} \left[-\,\frac78\right] \,. \end{equation}
 
  When $a\ll b_j\,$, Lukosz calculates an attractive pressure
 \begin{equation} P_a = -\,\frac{\pi^2}{240 a^4}\,, \end{equation}
 which implies by (\ref{Rayleigh}) Boyer's formula \cite{Boy}
 \begin {equation}
 \overline{P}_a = +\, \frac78\,  \frac{\pi^2}{240 a^4}
\label{inside}\end{equation}
 for the box with one permeable wall.
The external (long) part of the shaft has length $L-a \gg b_1=b_2=b$.
 For this limit, 
 Lukosz finds a repulsive pressure (involving Catalan's constant)
 \begin{equation} P = +\, \frac{0.915965}{24b^4}\,. \end{equation}
 Just as in \cite{HJKS}, the resulting force is inversely 
proportional to the cross-sectional area and is independent of~$L-a$, 
 so the corresponding energy term is proportional to~$L-a$.
 Therefore, application of (\ref{Rayleigh}) gives
 \begin{equation}
 \overline{P}_{L-a} = P_{L-a} =  +\, \frac{0.915965}{24b^4}
 \label{outside}\end{equation}
 (as ought to be the case, since the nature of the plate at the 
distant end of the long shaft ought to be irrelevant).
 To find the total force on the piston, we must reverse the sign of 
(\ref{outside}), add it to (\ref{inside}), and multiply by the 
area, $b^2$.
 The point is that the total force is positive if $a\ll b$;
 the long external part of the shaft has negligible effect in that case.
    
 On the other hand, for a cube Lukosz found that the perfectly 
conducting box was already repulsive.
 The formula (\ref{Rayleigh}) does not yield a simple factor 
 $-\frac12$ in that case, because the doubled box is no longer a 
cube.
 Nevertheless, the graph presented in \cite{HJV}  
 shows that $E_{2a}$ is closer to $\frac12 E_a$ than to $E_a\,$.
 We conclude that
 the permeable piston is attractive in the cubical 
configuration. 

\goodbreak


\begin{thebibliography}{00}

  \bibitem{Luk} W. Lukosz, Physica {\bf56}, 109 (1971).

 \bibitem{SS}
    N. F. Svaiter and B. F. Svaiter, J. Phys. A {\bf25}, 979 (1992). 

\bibitem{Cav} R. M. Cavalcanti, Phys. Rev. D {\bf69}, 065015 (2004).

\bibitem{HJKS} M. P. Hertzberg, R. L. Jaffe, M. Kardar, and A. 
Scardicchio, Phys. Rev. Lett. {\bf95}, 250402 (2005).

 \bibitem{Bar} G. Barton, Phys. Rev. D {\bf73}, 065018 (2006).

\bibitem{JR}  M. T. Jaekel and S.  Reynaud,
J. Phys. I (France) {\bf1}, 1395 (1991).  

\bibitem{perorb} M. Schaden and L. Spruch, Phys. Rev. A {\bf 58}, 935 
(1998).

\bibitem{JS}
 R. L. Jaffe and A. Scardicchio, Phys. Rev. Lett. {\bf 92}, 070402 (2004).

\bibitem{schad} M. Schaden,  Phys. Rev. A {\bf73}, 042102 (2006).

 \bibitem{Roth-1983} J.-P. Roth, in {\sl Th\'{e}orie du Potentiel,
 Proc. Colloq. J. Deny},
 G. Mokobodzki and D. Pinchon, eds. (Springer-Verlag, 1985).

 \bibitem{kottossmil} T. Kottos and U. Smilansky, Ann. Phys. (N.Y.) 
{\bf 274}, 76 (1999).

 \bibitem{Kuc} P. Kuchment, Waves Random Media {\bf14}, S107 
(2004).


\bibitem{gnutzmann} S. Gnutzmann and U. Smilansky,
  Advances In Physics {\bf 55}, 527 (2006).

 \bibitem{Kuc2002} P. Kuchment, Waves Random Media {\bf12}, R1
(2002).

\bibitem{systematics} S. A. Fulling, J. Phys. A {\bf36}, 6857 (2003).


 \bibitem{Wil} Justin H. Wilson, Undergraduate Research Fellow 
thesis, Texas A\&M University, in preparation.


    \bibitem{Fsb} S. A. Fulling, Contemp. Math. {\bf415}, 161 
(2006) (G. Berkolaiko et al., eds., {\sl Quantum Graphs and 
    Their Applications}).

 \bibitem{BM}
    B. Bellazini and M. Mintchev, J. Phys. A {\bf39}, 11101 (2006).

\bibitem{BHW} G. Berkolaiko, J. Harrison, and J. H. Wilson, in 
preparation.




\bibitem{kswire}   V. Kostrykin and R. Schrader, J. Phys. A {\bf32}, 595.

  \bibitem{ksinverse} V. Kostrykin and R. Schrader, Fortschritte der
Physik {\bf 48}, 703 (2000).


\bibitem{kostrykin-2007} V. Kostrykin, J. Potthoff, and R. Schrader,
  arXiv:math-ph/0701009 (unpublished).

\bibitem{ES} P. Exner and P. \v{S}eba, Rep. Math. Phys. {\bf28}, 7 (1989).

\bibitem{BF} J. D. Bondurant and S. A. Fulling, 
J. Phys. A {\bf38}, 1505 (2005).


 \bibitem{Boy} T. H. Boyer, Phys. Rev. A {\bf9}, 2078 (1974).

 \bibitem{AFG} D. T. Alves, C. Farina, and E. R. Granhen,
 Phys. Rev. A {\bf73}, 063818 (2006).

 \bibitem{IC} D. Iannuzzi and F. Capasso, Phys. Rev. Lett. {\bf91}, 
029101 (2003).

\bibitem{KKMRrep} O. Kenneth, I. Klich, A. Mann, and M. Revzen,
Phys. Rev. Lett. {\bf91}, 029102 (2003).

 \bibitem{SZL} C.-G. Shao, D.-L. Zheng, and J. Luo,
 Phys. Rev. A {\bf74}, 012103 (2006).

 \bibitem{Dow} J. S. Dowker, Nucl. Phys. B (Proc. Suppl.) 
 {\bf104}, 153 (2002).

\bibitem{HJV} S. Hacyan, R. J\'auregui, and C. Villarreal,
 Phys. Rev. A {\bf47}, 4204 (1993).

\end{thebibliography}
\end{document}